\documentclass[fleqn,usenatbib]{mnras}
\usepackage{newtxtext,newtxmath}
\usepackage[T1]{fontenc}

\usepackage{graphicx}
\usepackage{amsmath}
\usepackage[dvipsnames]{xcolor}
\usepackage[normalem]{ulem}
\graphicspath{{figs/}}
\usepackage{ae,aecompl}
\usepackage{color, colortbl}
\usepackage{wrapfig}
\usepackage{multicol}

\usepackage{color}
\usepackage{colortbl}

\title[Impact of PBHs on the mass function of UFDs]{The impact of primordial black holes on the stellar mass function of ultra-faint dwarf galaxies}

\author[Esser, De Rijcke, Tinyakov]{
Nicolas Esser$^{1}$\thanks{E-mail: nicolas.esser@ulb.be},
Sven De Rijcke$^{2}$\thanks{E-mail: sven.derijcke@ugent.be}, and
Peter Tinyakov$^{1}$\thanks{E-mail: petr.tiniakov@ulb.be}
\\
$^{1}$Service de Physique Th\'{e}orique, Universit\'{e} Libre
de Bruxelles (ULB), CP225 Boulevard du Triomphe, B-1050 Bruxelles,
Belgium \\
$^{2}$Department of Physics and Astronomy, Ghent University, Krijgslaan 281, 9000 Ghent, Belgium
}

\date{Accepted XXX. Received YYY; in original form ZZZ}

\pubyear{2023}

\begin{document}
\label{firstpage}
\pagerange{\pageref{firstpage}--\pageref{lastpage}}
\maketitle
\begin{abstract}
If primordial black holes (PBHs) constitute the dark matter (DM), stars forming in dark-matter dominated environments with low velocity dispersions, such as ultra-faint dwarf galaxies, may capture a black hole at birth. The capture probability is non-negligible for PBHs of masses around $10^{20}$~g, and increases with stellar mass. Moreover, infected stars are turned into virtually invisible black holes on cosmologically short time-scales. Hence, the number of observed massive main-sequence stars in ultra-faint dwarfs should be suppressed if the DM was made of asteroid-mass PBHs. This would impact the measured mass distribution of stars, making it top-light (i.e. depleted in the high-mass range). Using simulated data that mimic the present-day observational power of telescopes, we show that already existing measurements of the mass function of stars in local ultra-faint dwarfs could be used to constrain the fraction of DM composed of PBHs in the -- currently unconstrained -- mass range of $10^{19}$ -- $10^{21}$~g.

\end{abstract}


\section{Introduction}
Since the discovery of  gravitational waves from black hole mergers \citep{LV1,LV2}, 
the hypothesis of primordial black holes (PBHs) has attracted a lot of attention in the scientific community. These hypothetical black holes, first proposed in the late 1960s \citep{Zeldovich,Hawking}, could have been produced in the early Universe and potentially explain part, or the totality, of the dark matter (DM). However, many bounds have been placed on the fraction $f=\Omega_{\text{PBH}}/\Omega_{\text{DM}}$ of DM that can be made of PBHs, and currently only PBHs of masses between $\sim10^{17}$~g and $\sim10^{23}$~g can still make up the entirety of the DM \citep{Carr}. The black holes in this range of masses are particularly difficult to constrain as they are massive enough to avoid Hawking evaporation, but still of microscopic size, and of masses below the numerous lensing constraints that arise for heavier PBHs.

One way to constrain PBHs in this mass range is to use their capture by stars, which leads to the accretion of matter and conversion of the infected stars into sub-solar mass black holes on cosmologically short time-scales. Observationally, this effect would manifest itself as the disappearance of stars, since the resulting sub-solar mass
black holes are basically invisible. This process has been considered in application to neutron stars and white dwarfs by \cite{PBHSFNS,PBHdirectcapNS,PBHSFNS2} and to main-sequence stars by \cite{PBH1}. The advantage of the latter as
PBH detectors is that they themselves are much easier to observe.

The probability of PBH capture by a star is generally very small. It may only reach values of order one in exceptional environments, such as ultra-faint dwarf galaxies (UFDs). These systems are expected to be DM-rich, given that the concentration of DM-halos steadily increases with decreasing mass down to the resolution limit of current cosmological simulations \citep{2021MNRAS.506.4210I} and that their early supernova explosions are not sufficiently numerous to flatten the DM-cusp to a core \citep{2015MNRAS.453.2133B}. They also have low velocity dispersions which, combined with their high DM density, provides the most favourable conditions for capture. Requiring  that no more than half of the stars in the most promising UFD, Triangulum II, has been converted into black holes would limit the fraction of PBHs $f$ to at most 40\% \citep{PBH1}.

Based on the present-day observations, it is not easy to reliably constrain the fraction of stars that have been converted into black holes in a given dwarf galaxy over its history, as one would have to rely on modelling a UFD's evolution, which incurs large uncertainties. The purpose of the present paper is to demonstrate that constraining PBHs through star destruction is nevertheless feasible. The key observation is that the PBH capture, and hence the star destruction is significantly more efficient for more massive stars, and therefore overall this process suppresses the stellar mass function at larger masses. The stellar mass functions of many UFDs have been measured. We show that the already existing and near-future data may be sufficient, in quantity and quality, to constrain the impact of PBHs and, consequently, their abundance.

In this paper, we follow a proof-of-concept strategy and do not use real data. Instead, we generate a mock stellar sample which is similar in size, mass distribution, and other properties to the samples collected for real dwarf galaxies. We will refer to this sample as the “mock data”. By construction, this sample is unaffected by PBHs. We then use these mock data to estimate the constraining power of the stellar mass function with respect to possible effects of PBHs. To this end, we take as a model a general parametrization of the stellar mass function, either a broken power-law (BPL) or a log-normal (LN) distribution, each depending on a number of free parameters, and apply the effect of star destruction by PBHs. The fraction $f$ of PBHs is treated as a free parameter of the model. We use a Bayesian approach to fit the model to the mock data, and to determine the likely values of the model parameters, and in particular of the PBH-DM fraction $f$. With this approach, we find that it is possible to exclude the value $f=1$ for the PBH-free mock data with reasonable confidence. 

The rest of this paper is organized as follows. In Sec. \ref{sec:StarsInUFD}, we summarize the observational facts about dwarf galaxies relevant for this work.  In Sec. \ref{sec:stardestru}, we calculate the probability of the PBH capture by a star, as a function of the star mass. In Sec. \ref{sec:SMF}, we discuss the initial mass function (IMF) and construct a corresponding model of the present-day mass function (PDMF) that includes the effect of PBHs. In Sec. \ref{sec:Bayesian} we give the details of the mock data sample, and perform a Bayesian fit of the model to the mock data. In Sec. \ref{sec:frequentist} we cross-check our results with the frequentist approach. We will finally conclude in Sec. \ref{sec:conclusion}.

\section{DM and stars in dwarf galaxies}
\label{sec:StarsInUFD}

The UFDs are the most DM-dominated systems as inferred from the current observations \citep{Simon}, which makes them ideal candidates for constraining the abundance of PBHs. It was shown that the star formation in such systems was shut down at the time of reionization, around 1 billion years after the big bang \citep{reionization}. To a good approximation, the star population of such galaxies can thus be considered as isochrone, i.e. all the stars in a given UFD were born at the same time $\sim 12.8$ billion years ago. Over this time, stars more massive than $\sim 0.8 M_\odot$ would have evolved off the main sequence. Due to the faintness of their remnants, the impact of PBHs on the population of stars above this mass is complicated to measure, and we will concentrate on stars below this threshold in what follows.

Multiple photometric studies of low-mass stars in UFDs have been carried out in the last two decades. Using the stellar luminosities measured in different colour bands, the colour-magnitude diagram of these galaxies was obtained in order to study their chemical and dynamical properties \citep[see e.g.][]{Conn,Filion1,Baumgardt}. While all of these works involved star counting in the observed galaxies (typically containing between a few hundred and a few thousand individually resolved stars), a fit of the mass distribution of stars in these galaxies was only performed in \cite{feltzing}, \cite{Geha}, \cite{Gennaro1,Gennaro2} and \cite{Filion2}. Due to the limited sensitivity of telescopes, it was impossible for a long time to detect individual stars with masses below $\sim 0.5M_\odot$ in UFDs. However, recent observations were able to resolve stars with masses as low as $\sim 0.2M_\odot$ in some of these galaxies. Preliminary results from the James Webb Space Telescope even showed a resolution that reached $0.09M_\odot$ in the dwarf galaxy Draco II, located at a distance of $\sim20$ kpc \citep{jwst}. 

Other studies, as reviewed by \cite{Simon}, put bounds on different properties of these galaxies, such as their metallicity and velocity dispersion, both of which turn out to be extremely low in these systems. The masses of many UFDs were also estimated. From these measurements, one finds that the UFDs are completely dominated by DM, and obtains an estimate of the DM density. 
In the context of this paper, two parameters are of major interest: the DM density $\rho_{\text{DM}}$, and its 3-D velocity dispersion, $\sigma_v$\footnote{The 3-D velocity dispersion $\sigma_v$ used throughout this paper is related to the line-of-sight velocity dispersion $\sigma_\text{los}$ through $\sigma_v=\sqrt{3}\sigma_\text{los}$.}, which is assumed to be the same as the stellar velocity dispersion. Their combination $\rho_{\text{DM}}/\sigma_v^3$ controls the probability of the PBH capture by a star (see next section). When normalized to the reference values $\rho_\text{DM} = 100$~GeV/cm$^3$ and $\sigma_v =7$~km s$^{-1}$ it gives the “merit factor” $\eta$ introduced in \cite{PBH1},
\begin{equation}
\eta =    \dfrac{\rho_{\rm DM}}{100~
\text{GeV}/\text{cm}^3}\left(\dfrac{7~\text{km s}^{-1}}{\sqrt{2} \sigma_v}\right)^3. 
\label{eq:eta}
\end{equation}
Among currently observed UFDs, the galaxy Triangulum II was found to have the highest merit factor of $\eta=0.95$, while another four galaxies (Tucana III, Draco II, Segue I, Grus I) have merit factors exceeding 0.35. For some UFDs in this list, only upper limits on the velocity dispersion are available (and were used to compute $\eta$), indicating that their merit factors may be even larger than our current estimates. Future improvements in the measurement of $\sigma_v$ in dwarf galaxies would thus strengthen the results of the following sections.

\section{Star destruction by PBHs}
\label{sec:stardestru}

Main sequence stars capture PBHs most efficiently at the time of their formation. As a star is formed from a protostellar gas cloud, it may acquire a number of satellite PBHs that have orbits passing through the newly formed star. These PBHs then lose energy to dynamical friction every time they cross the star and, given enough time, finally end up captured inside the star. The only relevant orbits are the bound ones that repeatedly pass through the star since the energy lost in a single crossing is tiny. 

The capture is an uncorrelated random process, and therefore the number of PBHs captured by a given star follows a Poisson distribution characterized by the mean captured PBH number, $\bar N = \bar{M}_{\text{cap}}/m$, $m$ being the mass of the PBH and $\bar{M}_{\text{cap}}$ the mean PBH mass captured by the star. If a star captures at least one PBH, it  will inevitably be destroyed; the probability of survival is thus $P_S=\exp(-\bar{N})=\exp(-\bar{M}_{\text{cap}}/m)$. 

The mean captured mass $\bar{M}_{\text{cap}}$ depends on both the PBH mass $m$ and the star mass $M$ (all other relevant stellar parameters are evaluated as functions of $M$), as well as on the local DM density $\rho_{\rm DM}$, the velocity dispersion $\sigma_v$, and on the fraction of PBH $f$ in the total amount of DM. In a wide mass range $m=10^{19}$ -- $10^{21}$~g, which we will focus on in what follows, the dependence on most of these parameters is trivial. The captured mass is obviously proportional to the PBH fraction $f$. It is also proportional to the combination $\rho_{\rm DM}/\sigma_v^3$ which is, up to normalization, the merit factor $\eta$ (\ref{eq:eta}). Indeed, only the low-velocity tail of the PBH distribution is prone to capture, with the upper limit being set by the parameters of the protostellar cloud. The amount of DM in this region of the phase space is proportional to $\rho_{\rm DM}/\sigma_v^3$. 
Finally, as has been shown by numerical simulations in \citet[their fig. 1]{PBH1}, in this mass range it is also proportional to the PBH mass $m$. Therefore, one can write the mean captured number $\bar N$ as 
\begin{equation}
\bar N = \bar{M}_{\text{cap}}/m = f \eta\, \nu(M)
\end{equation}
where $\nu(M)$ is the mean captured number $\bar{M}_{\text{cap}}/m$ calculated for the reference values of the DM parameters $\eta=1$ and $f=1$. The function $\nu(M)$ only depends on the star mass $M$. The star survival probability can thus be written as 
\begin{equation}
    P_S(f,M) = \exp[-f\eta\,\nu(M)].
    \label{eq:surviprob}
\end{equation}

\begin{figure}
\includegraphics[width=1.\columnwidth]{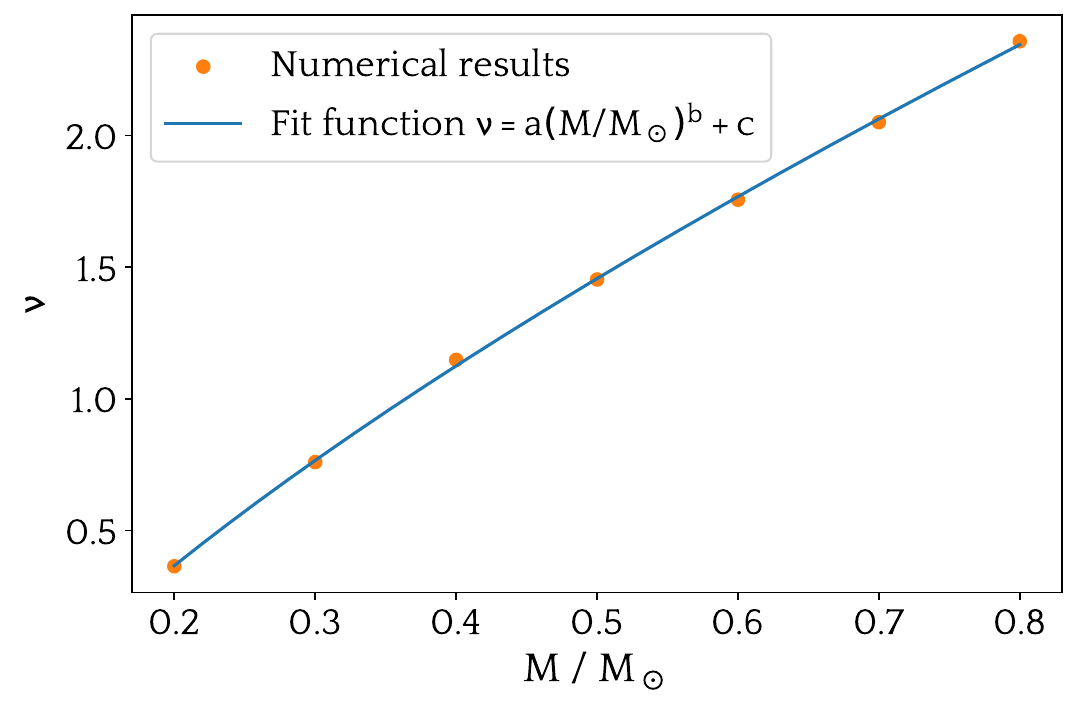}
\caption{\label{fig:capM}
Mean captured PBH number $\nu$ for the reference values $\eta=f=1$ as a function of the stellar mass $M$. The statistical uncertainties of
the simulations are too small to be visible on the plot. The blue line shows the fit with the function \eqref{eq:fit}.}
\end{figure}

The function $\nu(M)$ plays a key role in our analysis. We show this function in Fig.~\ref{fig:capM}, computed numerically for $12.8$~Gyr old stars with stellar masses in the range $M\in[0.2,0.8]M_\odot$ using the method described in \cite{PBH1}. There, the mean mass $\bar{M}_{\text{cap}}$ of PBHs captured by a star at birth was computed. The time spent by the PBH to sink towards the star and acquire a trajectory which is enclosed in the star was taken into account in this calculation. The possibility that the PBH could get pushed out of its trajectory by other astrophysical objects in the vicinity of the star was also considered. Finally, it was checked that, once the PBH is completely trapped inside, it destroys the star in a finite time.
Note that $\nu(M)$ was computed for uniform DM and stellar distributions within the dwarf galaxy. We however checked that the results remain similar in case of non-uniform profiles. Since $\nu(M)$ is independent of the PBH mass in the interval $[10^{19},10^{21}]$~g, these results apply not only to the case where all PBHs have the same mass, but also to any non-monochromatic population of black holes whose masses lie between these two values.

For the sake of numerical efficiency, we fitted the results of the computations with an analytical function 
\begin{equation}
\label{eq:fit}
\nu(M) = a\left(M/M_\odot\right)^b+c
\end{equation}
with the parameters $a=3.76$, $b=0.685$ and $c=-0.883$. This fit is represented by the blue line in Fig.~\ref{fig:capM}.

\section{Impact of PBHs on the stellar mass function}
\label{sec:SMF}
\subsection{Present-day mass function model}
According to the isochrone hypothesis, the stellar population of UFD galaxies is born during one initial star-formation burst, and is unchanged after that. Hence, in the absence of PBHs, and for stars below $\sim0.8M_\odot$ that did not turn into compact objects yet, the present-day observed mass distribution is the same as the initial mass function (IMF). 

The presence of PBHs changes this picture, as a fraction of the stars in the galaxy are destroyed at or after the initial star-formation burst. The effect of PBHs on the present-day stellar mass function can therefore be easily calculated from the IMF and the probability of PBH capture.

The IMF, i.e. the mass distribution of stars at birth, has been one of the leading topics of research in stellar astrophysics for several decades. While there exist some standard and universal fits to the IMF \citep{Kroupa,Chabrier}, there likewise exists ample evidence for IMF variations in dwarf galaxies \citep{Geha,Gennaro1,Gennaro2}. 

More generally, the IMF can be modelled with a broken power-law (BPL) distribution
\begin{equation}
\dfrac{dN(\alpha_1,\alpha_2,M)}{dM}\propto
\begin{cases}
&M^{\alpha_1} \text{ for }0.2M_\odot\le M<0.5M_\odot\\
&kM^{\alpha_2} \text{ for }M\ge0.5M_\odot,
\end{cases}
\label{eq:powerlaw}
\end{equation}
with the continuity constant $k=(0.5M_\odot)^{(\alpha_1-\alpha_2)}$ and $\alpha_{1,2}$ the two power-law exponents. Here we have adopted $M = 0.5M_\odot$ for the break mass between the two power-laws, as it is the commonly used value, and current observations (cf. e.g. \citealt{Gennaro2}) show little deviation from this number. Note that for $\alpha_1=-1.3$ and $\alpha_2=-2.3$, the \cite{Kroupa} IMF is recovered. In case one solely considers stars of masses above $0.5M_\odot$, the only relevant parameter left is $\alpha_2$. We will call this particular case the single power-law (SPL) distribution.

Another model which is often used to describe the IMF is the log-normal (LN) distribution,
\begin{equation}
\dfrac{dN(M_c,\sigma_{\text{LN}},M)}{dM}\propto \dfrac{1}{M}\exp\left(-\dfrac{\left(\log_{10}(M/M_c)\right)^2}{2\sigma_{\text{LN}}^2}\right),
    \label{eq:LN}
\end{equation}
where the two free parameters are the characteristic mass $M_c$ and the width $\sigma_{\text{LN}}$ of the distribution. The (single-star) \cite{Chabrier} IMF is recovered for $M_c=0.08M_\odot$ and $\sigma_{\text{LN}}=0.69$.

\begin{figure}
\includegraphics[width=1.\columnwidth]{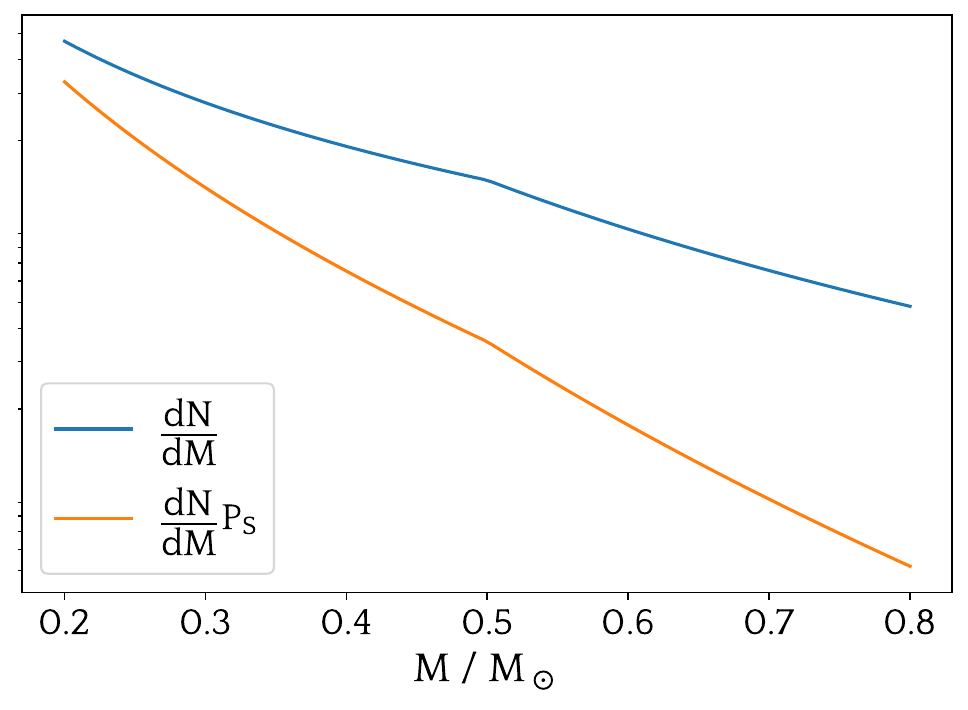}
\caption{\label{fig:IMFvsPDMF} Kroupa IMF (upper line, blue) and associated PDMF (lower line, orange) after the destruction of a fraction of the stars by PBHs. A logarithmic scale is used on the vertical axis. The normalization of the IMF, and hence the scale on the vertical axis, is arbitrary.}
\end{figure}

With the isochrone hypothesis in mind, one may write the normalized present-day mass function (PDMF) of the stars in a given UFD, including the destruction by PBHs, as follows
\begin{equation}
f_{\text{PDMF}}(f,\theta,M) =C \dfrac{dN}{dM} P_S
\label{eq:PDMF}
\end{equation}
with $C$ the normalization factor  
\begin{equation}
    C^{-1}(f,\theta)=\int_{M_{\text{min}}}^{M_{\text{max}}}\dfrac{dN}{dM} P_S dM
\end{equation}
such that the PDMF is normalized to $1$.
Here $\theta$ stands for the parameters of the chosen IMF which are $(\alpha_1$, $\alpha_2)$ for the BPL; $(M_c$, $\sigma_{\text{LN}})$ for the LN; and $\alpha_2$ for the SPL. The factor $P_S(f,M)$ is the survival probability of a star against its destruction by PBHs over the last $12.8$ billion years. Its presence in eq.~\eqref{eq:PDMF} quantifies the impact of PBHs on the stellar mass function. Note that without the isochrone approximation, $P_S$ needs to be computed individually for each star, taking into account its age. To better visualize the impact of PBHs on the mass function, in Fig.~\ref{fig:IMFvsPDMF} we show, as an example, the Kroupa IMF (BPL with $\alpha_1=-1.3$ and $\alpha_2=-2.3$) and the resulting PDMF after destruction by PBHs. More massive stars are destroyed with higher probability, which translates into a PDMF with a steeper slope than the IMF.

\subsection{Remark on binary systems}
The comparison of model mass functions with observations is in general complicated by the presence of binary systems, which in most cases cannot be resolved. These unresolved systems appear as a single bright object, and thus tend to make the observed mass function more top-heavy than what the real stellar mass distribution is. In most observational mass function studies, these unresolved systems are taken into account through one extra parameter, the binary fraction.

The PBH capture by stars in binary stellar systems has not yet been considered in the literature. It involves three-body dynamics, which may modify the capture rate. On the one hand, the presence of a companion may deviate the trajectories that are passing through the star and prevent the capture. On the other hand, the trajectories not passing through the star may start doing so as a result of the perturbation by the companion. The net effect of a companion star, if any, is therefore unclear. 

One can take a conservative approach, assuming that all the trajectories that are sufficiently perturbed are not captured, and calculate the fraction of the PBH trajectories that are not impacted by the presence of companion stars. Using the database of trajectories from the previous simulations of \cite{PBH1}, we can determine the probability distribution of the PBH apastrons in their orbits around the star. Assuming that the typical distance $d$ between the partners in a binary is distributed as $1/d$ for $d\in[15,10^5]$au \citep{binarydistrib} --- which again is a conservative assumption for newly formed stars around which the PBHs orbit --- one can use the 3D analogue of equation~(6) of \cite{PBH1} to find the fraction of PBH orbits that are captured even in the presence of a companion. We found that this fraction varies between 15\% and 65\% depending on the combination of stellar masses and on the PBH mass, being generally larger for less massive PBHs. 

To give an example, in dwarf galaxies Coma Berenices and Bo\"otes I, for the star mass range similar to what we are using, the binary fractions were found to be $\sim0.35$ and $\sim0.7$, respectively \citep{Gennaro2,Filion2}, with  $\sim 50$\% uncertainties. 
So, assume for simplicity that half of the stars are in binaries that are not resolved and appear as single objects. Then of all observed 'stars', 2/3 are actual stars and capture the PBH with the rate set by Fig.~\ref{fig:capM}, while 1/3 are unresolved binary systems that capture the PBH at a rate reduced by, say, 50\% on the basis of the previous discussion. Clearly, the resulting overall loss in the capture rate of $\sim 15\%$ is a correction that for the purposes of this paper can be neglected. We will thus disregard the issue of binary stars in what follows.

\section{Bayesian analysis of the PDMF model}
\label{sec:Bayesian}

\subsection{Mock data}
\label{sec:mockdata}

To mimic the real data, we generate a mock sample of stars (referred to as “mock data” in what follows) with parameters that resemble the observed ones. We will later run it through the analysis as if it were real data. 
Since we aim at constraining PBHs, we do not include any PBH effects in this sample. 

Specifically, the mock isochrone stellar population is generated using the MIST \citep{mist0,mist1} isochrone models and the ArtPop  {\sc python} package \citep{artpop}. All the stars in the sample have a metallicity $[\text{Fe/H}]=-2.2$ and an age of $12.8$ billion years, which are values typical of UFDs \citep{Simon}. They follow the \citet{Kroupa} IMF, i.e. the distribution \eqref{eq:powerlaw} with $\alpha_1=-1.3$ and $\alpha_2=-2.3$. 

Based on the review of the existing measurements of individual stars in local dwarf galaxies, cf. Sec. \ref{sec:StarsInUFD}, we choose the mock sample to contain $N=1000$ stars with masses between $M_{\text{min}}=0.2M_\odot$ and $M_{\text{max}}=0.79M_\odot$. The number of stars in the mock sample is of the same order as what has been measured in different UFDs. The upper bound $M_{\text{max}}=0.79M_\odot$ is the mass of the most massive stars, within the MIST model, that 
are still on
the main sequence in a 12.8~Gyr old galaxy. The lower bound of $0.2M_\odot$ corresponds to the smallest stellar mass that can currently be resolved in dwarf galaxies. One may expect that recent and future telescopes, such as the James Webb Space Telescope, will resolve stars of even lower masses in UFDs in the future \citep{jwst}.  

\begin{figure}
\includegraphics[width=1.\columnwidth]{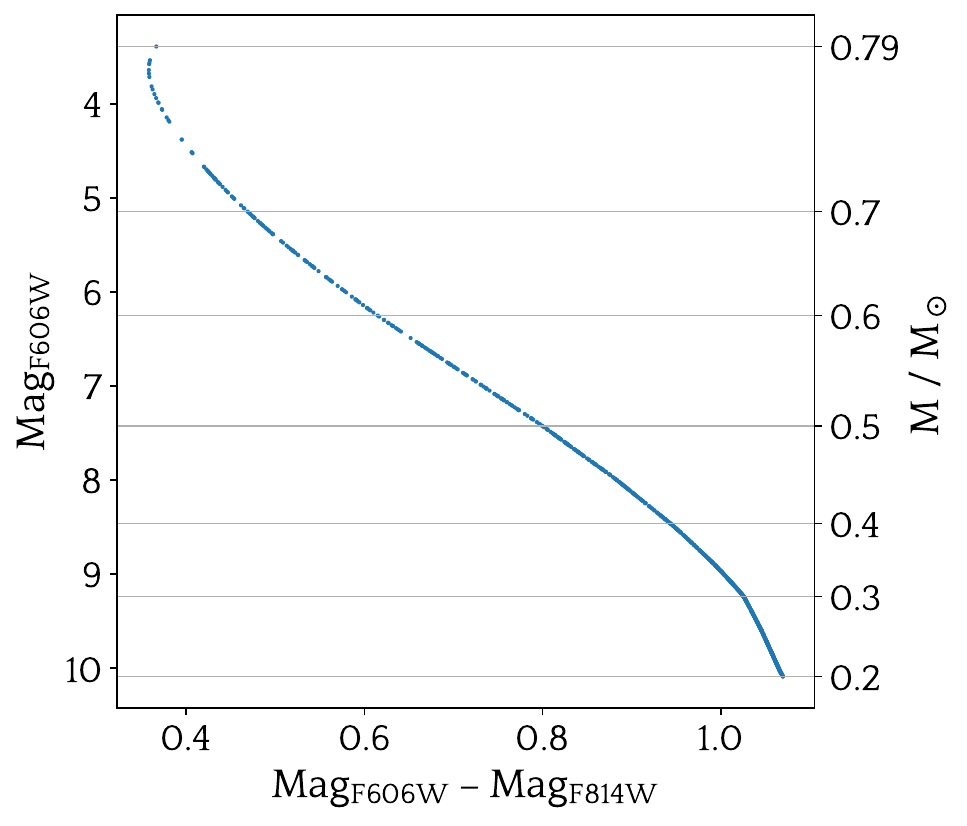}
\caption{\label{fig:CMD}
Colour-magnitude diagram of the mock data. The absolute magnitudes (in the F606W and F814W filters described in the text) are displayed on the left vertical axis, while the stellar masses are shown on the right vertical axis. They are a function of the colours of the stars (i.e. the magnitude difference between the two bands), displayed on the horizontal axis.}
\end{figure}

The stellar population of the mock sample is represented on a colour-magnitude diagram, Fig.~\ref{fig:CMD}. The absolute magnitude is displayed on the left vertical axis, while the associated masses are displayed on the right. The F606W and F814W filters [from the Advanced Camera for Surveys (ACS) on board of the \textit{Hubble Space Telescope (HST)}] are used to describe the luminosity and colour of the stars. Note that for simplicity, the population is considered at a distance of $10$ pc, so that the apparent and absolute magnitudes are the same. Reddening and extinction are neglected. These mock data are obviously idealized, and real observations would show a more scattered distribution around the isochrone curve. To assess the impact of noise on the data, we have also generated a -- more realistic -- scattered mass sample and performed the analysis described in Appendix~\ref{appendix}. The results were very similar to those obtained with the idealized data depicted in Fig.~\ref{fig:CMD}.

\subsection{Likelihood}
\label{sec:likelihood}

Given the current observational capabilities to resolve stars in UFDs as encoded in the mock data, we want to evaluate if the modifications to the stellar mass function of these dwarf galaxies due to PBHs are detectable. For this, we use a Bayesian approach. Given a sample of $N$ stars with individual masses $M_i$, either observed or simulated, one may define a binless likelihood $L$ that quantifies how well this set follows the distribution (\ref{eq:PDMF}) by taking the product of $f_{\text{PDMF}}(f,\theta,M)$ over all stars in the sample. Its logarithm 
\begin{align}
&\log L(f,\theta|\{M_i\}) =\sum_{i=1}^N \log(f_{\text{PDMF}}(f,\theta,M_i))
\label{eq:logL}
\end{align}
\begin{align*}
&=N\log(C(f,\theta))+\sum_{i=1}^N\log\left({\dfrac{dN(\theta,M_i)}{dM}}\right)+\sum_{i=1}^N\log\left(P_S(f,M_i)\right),
\end{align*}
is a standard quantity used for parameter estimation in statistical analysis. 

As defined above, the log-likelihood function depends on a number of parameters: the fraction of PBHs in the total amount of the DM $f$ and the parameters of the stellar IMF ($(\alpha_1$, $\alpha_2)$ for BPL; $(M_c$, $\sigma_{\text{LN}})$ for LN; $\alpha_2$ for SPL). Our primary interest is the estimation of the fraction $f$. The other, IMF-dependent parameters are considered as nuisance parameters. As for the merit factor, it was fixed for simplicity to the highest measured value of $\eta=0.95$. Note however that from eq.~\eqref{eq:surviprob} $\eta$ and $f$ are completely degenerate parameters. Hence, one can easily rescale $\eta$ to any value by adapting accordingly the result obtained for $f$.

\subsection{Bayesian analysis}
\label{sec:BayesAnalysis}

We compare our mock data sample to three IMF models: BPL, LN and SPL. In each of these models, we include the star destruction by PBHs, considering their fractional abundance $f$ as a free parameter to be determined from the fit together with other parameters of the model. The goal is to see how strongly the value of $f$ is constrained by the mock data which resembles the existing observations and, by construction, has no PBH effects in it.  

To estimate the model parameters, including $f$,  we use the likelihood $L$ defined in \eqref{eq:logL}. 
For the BPL and LN cases, the log-likelihood depends on three parameters: the fraction $f$ and the two IMF parameters: $\alpha_1$ and $\alpha_2$ for the BPL distribution, and $M_c$ and $\sigma_{\text{LN}}$ for the LN one. For the SPL case, the likelihood depends only on $f$ and $\alpha_2$.
In the latter case, the range of stellar masses is restricted to $M\geq 0.5M_\odot$ in order to mimic the data in which stars below $0.5M_\odot$ are not resolved. Keeping only the stars above this threshold reduces the size of our mock data sample for the SPL case from $N=1000$ to $N=233$ stars.

Based on the existing measurements of the IMFs of ultra-faint dwarfs galaxies with very low merit factors $\eta$, in which the destruction of stars by PBHs is expected to be negligible \citep{feltzing,Geha,Gennaro1,Gennaro2,Filion2}, we constrained the power-law exponents $\alpha_{1,2}$ to values $\in[-2.6,-0.8]$ and the LN parameters $M_c\in[0.08,0.6]$ and $\sigma_{\text{LN}}\in[0.5,0.7]$, as there are very few measured parameters outside these bounds. Note that these bounds are conservative regarding the presence of unresolved binary systems, as they encompass some observations which fit the single-star mass function and others that fit the system mass function. On the other hand, the fraction $f$ is by definition in the range $[0,1]$. Using uniform priors $\Pi(f,\theta)$ for all the parameters within their bounds, the posterior probability density function (pdf) can be computed for every IMF model as
\begin{equation}
    \text{pdf}(\{M_i\}|f,\theta)=\dfrac{\Pi(f,\theta)L(f,\theta|\{M_i\})}{\int_{f',\theta'}\Pi(f',\theta') L(f',\theta'|\{M_i\})},
    \label{pdf}
\end{equation}
where $\Pi$ is in this case a flat window function for each parameter and $\theta$ stands for all the parameters of the chosen IMF except $f$. 

One can see from eq.~\eqref{pdf} that the pdf is equal to the likelihood, to within a normalization factor. In order to avoid calculating this factor, which is computationally expensive for a sample size of $N=1000$ stars, we mapped the pdf numerically by rejection sampling. The method is straightforward: a random point $(f_j,\theta_j)$ is generated within the prior boundaries with uniform probability. A random uniform number $x_j \in[0,\max(L)]$ is also drawn. If $L(f_j,\theta_j|\{M_i\})>x_j$, the point is kept. It is discarded otherwise. The process is repeated until $n=10^6$ points are obtained. 
The distribution of these points in the parameter space then allows one to map the posterior pdf.

\begin{figure*}
\includegraphics[width=1.\columnwidth]{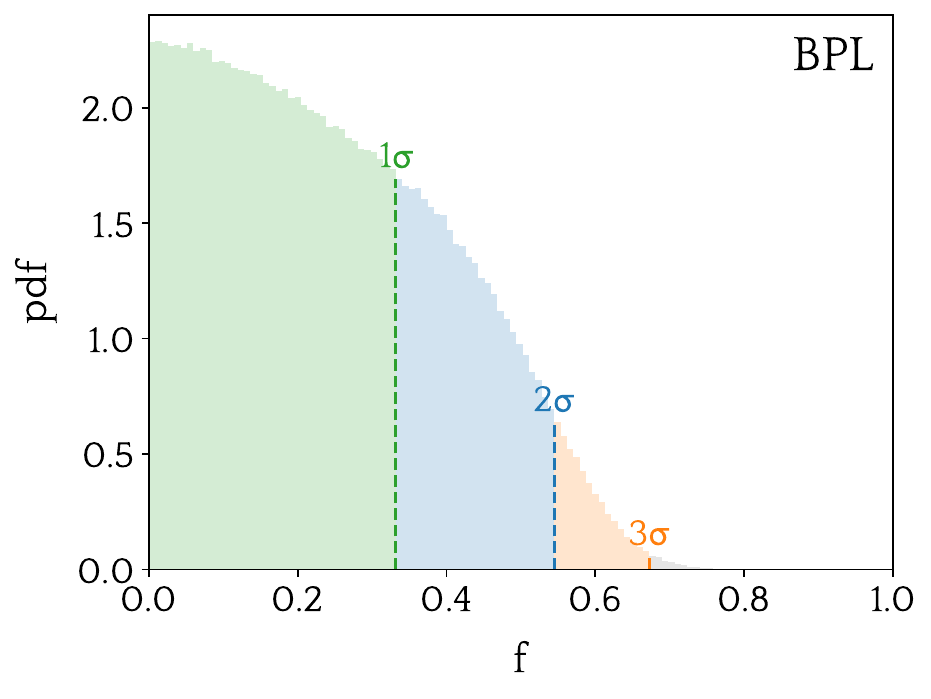}
\includegraphics[width=1.\columnwidth]{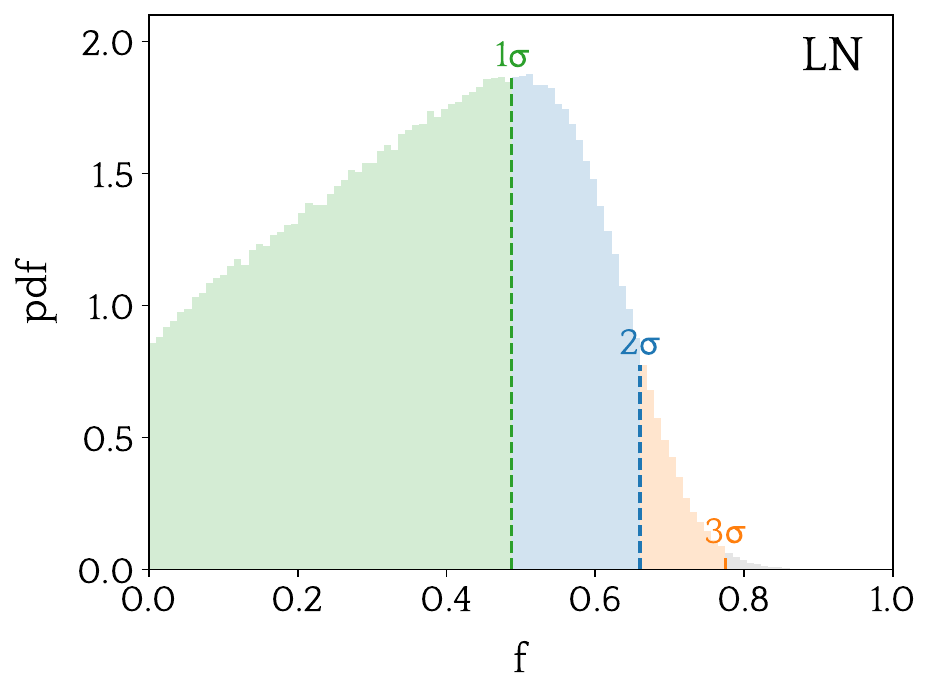}
\includegraphics[width=1.\columnwidth]{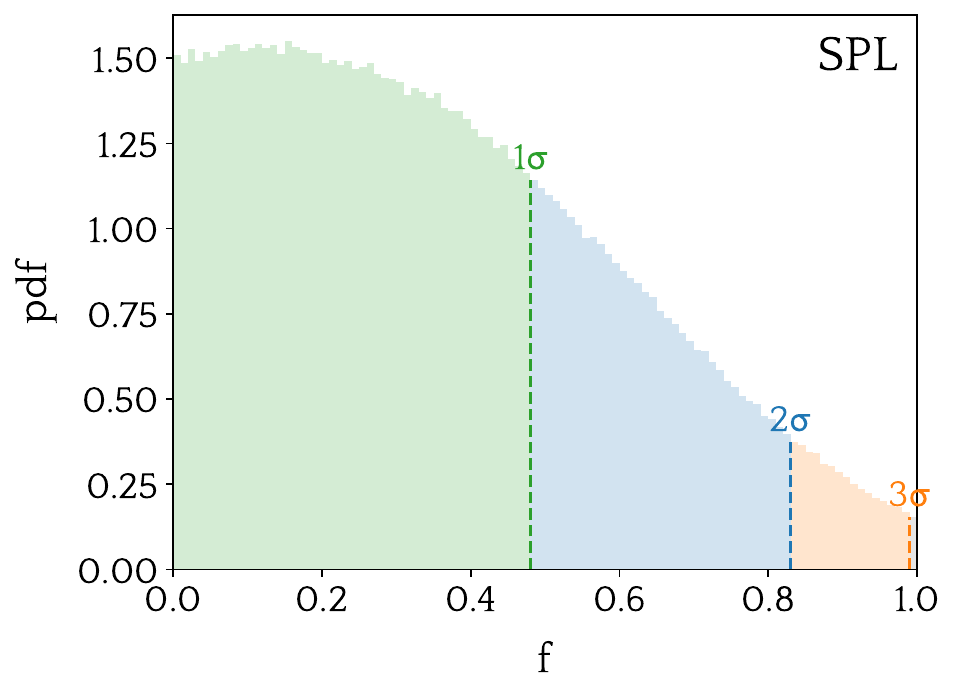}
\caption{\label{fig:BPLandLNandSPLintegrated}
Marginalized pdf of the PBH fraction $f$ for the broken power-law (top left panel), log-normal (top right panel) and single power-law (bottom panel) cases. The left (green) region corresponds to the values that are within one-sided $1\sigma$, the middle (blue) region corresponds to points between $1\sigma$ and $2\sigma$ and the right (orange) region corresponds to points between $2\sigma$ and $3\sigma$.}
\end{figure*}

In the case where the values of the (nuisance) parameters $\theta$ are unknown, one can marginalize the pdf distribution with respect to these parameters in order to get the pdf of $f$ alone, $\text{pdf}_{\text{marg}}(\{M_i\}|f)=\int_{\theta'}\text{pdf}(\{M_i\}|f,\theta')d\theta'$. This corresponds to drawing a histogram of the distribution of the points $f_j$ previously obtained, regardless of the values of the other coordinates $\theta_j$ of these points. For the BPL, LN and SPL cases respectively, these histograms are shown on the top left, top right, and bottom panels of Fig.~\ref{fig:BPLandLNandSPLintegrated}.

Using one-sided confidence levels, one can read from these histograms that for the BPL and LN initial mass functions, the fractions $f\gtrsim0.65$ and $f\gtrsim0.75$ are respectively excluded with a confidence of $3\sigma$. The hypothesis of all DM being made of PBHs, $f=1$, can be firmly excluded in these two cases. However, in the SPL case, $f\simeq1$ is only barely $3\sigma$ excluded.

The first thing we would like to point out regarding these results is that the exclusion regions resulting from the BPL and LN cases are similar, even though the pdfs of these two distributions are different. This is consistent with the fact that these mass functions are very close to one another. The exclusion resulting from the SPL case is less stringent, expressing the fact that the latter analysis only considered stars above $0.5M_\odot$, which represents $\sim1/4$ of the full sample of $1000$ stars.

Secondly, we recall that the sensitivity depends on the chosen priors for the nuisance parameters. More stringent constraints would arise from narrower window functions on $\theta$. The chosen windows used to produce these results are benchmark values based on observational data, as we lack consistent theoretical models for the mass function of stars in dwarf galaxies. Future improvements on the modelling of these parameters would impact the sensitivity obtained here.

Finally, we would like to emphasize once more that these results are specific to the generated mock data set. Another data sample, real or computer-generated, could give results that fluctuate from what has been obtained here. We however analysed several other randomly generated mock data sets and found that, for $N=1000$, $f=1$ was always excluded at more than $3\sigma$ in the BPL and LN cases, and at more than $2\sigma$ in the SPL case.

\section{Consistency check : frequentist analysis}
\label{sec:frequentist}
The goal of this section is to check the robustness of the results obtained with the Bayesian analysis by comparing them with results from a frequentist approach. For simplicity, we will focus on the BPL model (eq.~\eqref{eq:powerlaw}) for the IMF and fix $\alpha_1=-1.3$ (as in the Kroupa IMF) in what follows. Since PBHs destroy more high-mass than low-mass main-sequence stars, the range of stellar masses that is most impacted by PBHs is the one above $0.5M_\odot$, which is the reason we are keeping $\alpha_2$ free. The full pdf distribution of the PDMF then only depends on $f$ and $\alpha_2$ and can be represented easily on a plot. 

\begin{figure}
\includegraphics[width=1.\columnwidth]{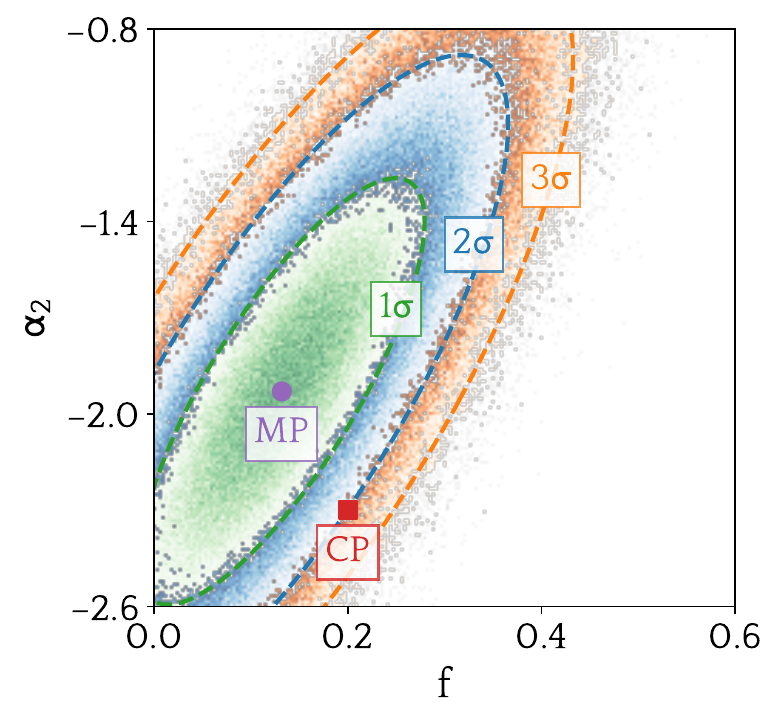}
\caption{\label{fig:2DBayesian}
pdf for the BPL case with $\alpha_1=-1.3$ fixed. The purple dot corresponds to the location of the MP of the real sample, while the red square is an arbitrary point used for comparison (CP). The innermost (green) region corresponds to the values that are within $1\sigma$, the intermediate (blue) region corresponds to points between $1\sigma$ and $2\sigma$ and the outermost (orange) region corresponds to points between $2\sigma$ and $3\sigma$.}
\end{figure}

First, using the Bayesian approach described in the previous section, we mapped the corresponding  pdf, cf. Fig.~\ref{fig:2DBayesian}. The maximum of the pdf (MP) located at $(f=0.132,\alpha_2=-1.93)$ is indicated with a purple dot. We then chose a specific point (call it CP, for “comparison point”), represented with a red square, which corresponds to the parameters $(f=0.2,\alpha_2=-2.3)$. The choice of this point is arbitrary. However, note that it is located on the boundary of the $2\sigma$-exclusion region of this pdf.

Now, in the frequentist approach, we generate many (ideally, an infinite number) trials of a specific model, with fixed values of the parameters $(f,\alpha_2)$ corresponding to the comparison point, and check how often the MP of the real sample (in our case, the mock data) is recovered given this model. In particular, choosing the BPL model with values of the parameters fixed to the ones of the comparison point, i.e. $f=0.2$ and $\alpha_2=-2.3$, we expect the maximum of the pdf of the real sample to fall around the $2\sigma$ boundary of the new maxima distribution, in order for the Bayesian and frequentist approaches to be consistent with each other.

\begin{figure}
\includegraphics[width=1.\columnwidth]{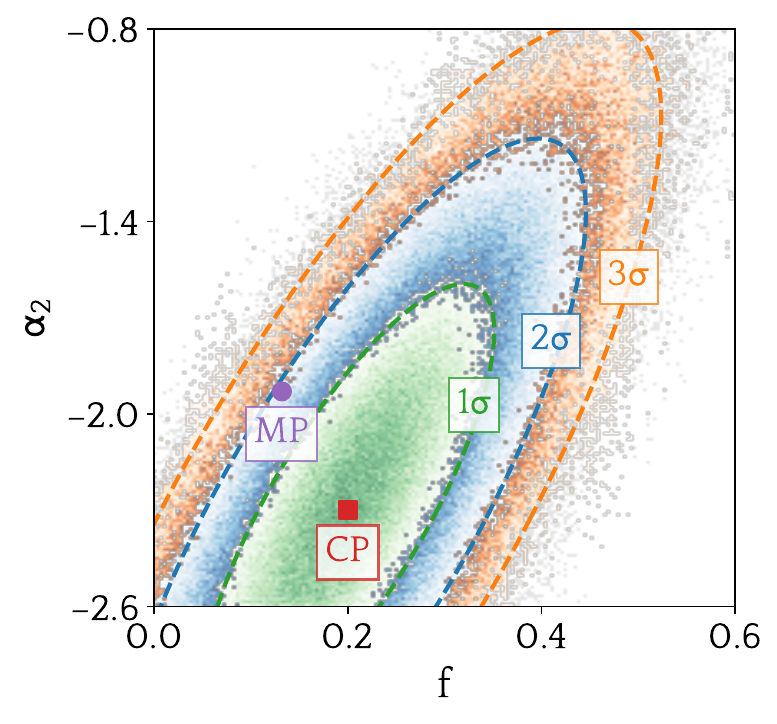}
\caption{\label{fig:2Dfrequentist}
Distribution of the maxima of the likelihoods for the BPL case with $\alpha_1=-1.3$ fixed. The fake populations were generated with $f=0.2$ and $\alpha_2=-2.3$. The purple dot corresponds to the location of the MP of the real sample, while the red square is an arbitrary point used for comparison (CP). The innermost (green) region corresponds to the values that are within $1\sigma$, the intermediate (blue) region corresponds to points between $1\sigma$ and $2\sigma$ and the outermost (orange) region corresponds to points between $2\sigma$ and $3\sigma$.}
\end{figure}

To verify that this is the case, we generated $10^6$ fake populations, using the ArtPop package with the hypotheses described in Sec. \ref{sec:mockdata} and with specific values of the input parameters $(f=0.2,\alpha_1=-1.3,\alpha_2=-2.3)$. We computed the maximum likelihood for each of these populations. The density distribution of these maxima in the parameter space is shown on Fig. \ref{fig:2Dfrequentist}.

We see that, obviously, the point CP is now the maximum likelihood of this new distribution. More importantly, the point MP falls around the $2\sigma$ border, as expected. The slight offset from this boundary is due to the finite number of stars in the populations, and the finite number of fake populations that were generated in the frequentist case.

We have repeated this whole exercise with other IMF models and specific parameter values, always finding the consistency between Bayesian and frequentist analyses. Hence, our results are robust and independent of the statistical approach employed.

\section{Conclusion}
\label{sec:conclusion}
The main conclusion from this work is that it is possible to constrain PBHs in the mass range $10^{19}$ -- $10^{21}$~g using already existing photometric observations of UFDs. The constraints, perhaps marginal with the existing data, could be improved in the future with observations from higher resolution telescopes and a better understanding of the dynamical properties of UFDs.

In order to demonstrate this, we first computed, based on the results of \cite{PBH1}, the probability of stars of different masses being destroyed by PBHs in the last $12.8$~Gyr. This probability depends on the fraction of DM made of PBHs $f$, as well as the dynamical properties of the DM in a given UFD. The key observation is that this probability increases with the mass of the star. 

Using the isochrone hypothesis and the fact that the initial and present-day mass function of stars below $\sim0.8M_\odot$ should be equal in the absence of PBHs, we can use the existing observations of low merit factor UFDs in which the effect of PBHs is negligible to parametrize the IMF. Assuming there is no strong dependence of the IMF on the merit factor, we then modeled the impact of PBH destruction on the PDMF. Because of preferential destruction of heavier stars, the resulting PDMF becomes more top-light, i.e. is suppressed on the high-mass end. In this way, we computed the model PDMFs corresponding to three different commonly used parametrizations of the IMFs. Note that future improvements in the understanding and measurement of the IMF of UFDs will enable one to obtain finer PDMF models and further strengthen the results.

We then generated a mock data set, using the ArtPop stellar population generation package, which mimics current observations of UFDs. This mock data set was not affected by PBHs. We performed a Bayesian analysis of this data set, given the PDMF model computed previously, in order to determine the best-fitting values of the PDMF model parameters, including $f$. 
We found that the value $f=1$ (all the DM is in the form of PBH) could be excluded at more than $3\sigma$ for the broken power-law and log-normal IMFs, and at more than $2\sigma$ for the single power-law with a restrained mass range. Finally, we checked, by generating fake populations using the ArtPop package, that the Bayesian results are consistent with a frequentist approach.

In this paper we have improved on the results of \cite{PBH1} in two respects. First, we have identified a more robust observable sensitive to the effect of PBH --- the UFD stellar mass function --- that is less prone to astrophysical uncertainties than a mere fraction of destroyed stars. Secondly, we have developed the Bayesian approach to quantify the statistical significance of the constraints on the PBH fraction. 

As a final remark, we note that basically the same approach can be used to discover PBHs rather than constrain their fraction. When applied to the real observations, one may in principle find that the value $f=0$ is not compatible with the data, which would be an indication of the PBH existence and a motivation for dedicated searches for these hypothetical objects. 

\section*{Acknowledgements}
SDR acknowledges support from grant Segal ANR-19-CE31-0017 of the French Agence Nationale de la Recherche.
NE is a FRIA grantee of the Fonds de la
Recherche Scientifique – FNRS.
PT is supported in part by the Institut Interuniversitaire des Sciences Nucl\'eaires (IISN) Grant No. 4.4503.15. 
Computational resources have been provided by the Consortium des \'Equipements de Calcul Intensif (C\'ECI), funded by the Fonds de la Recherche Scientifique de Belgique (F.R.S.-FNRS) under Grant No. 2.5020.11 and by the Walloon Region.

\section*{Data availability}
The data used in this article were generated numerically using the MIST \citep{mist0,mist1} isochrone models and the ArtPop {\sc python} package \citep{artpop}. The data will be shared on request to the authors.

\bibliographystyle{mnras}
\bibliography{bibli}

\appendix

\section{Scattered sample analysis}

\begin{figure}
\includegraphics[width=1.\columnwidth]{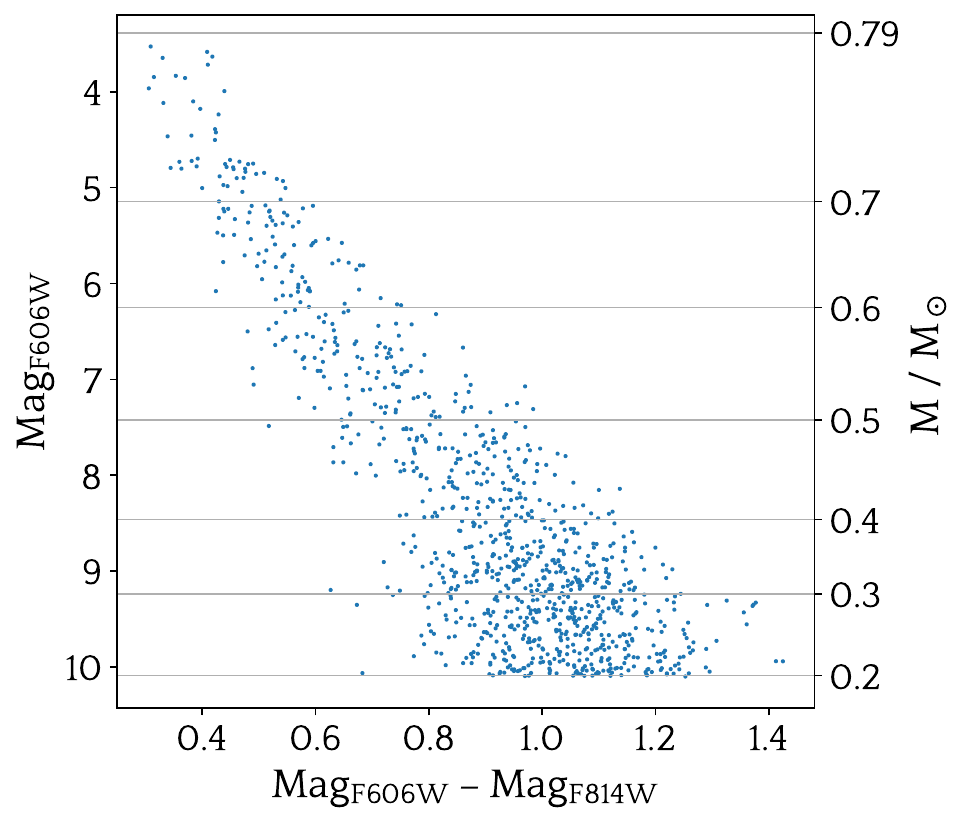}
\caption{\label{fig:CMD_UncertPop}
Colour-magnitude diagram of the noisy mock data. The absolute magnitudes (in the F606W and F814W filters described in the text) are displayed on the left vertical axis, while the stellar masses are shown on the right vertical axis. They are a function of the colours of the stars, displayed on the horizontal axis.}
\end{figure}

In reality, when trying to infer the mass function of an isochrone stellar population, multiple sources of uncertainties have to be taken into account. Apart from unresolved binaries, which we already mentioned in the main text, the stars do not have exactly the same age or metallicity. Hence, they do not follow exactly the same isochrone curve, which results in some extra scatter on the colour-magnitude diagram. Furthermore, there is for each star an associated photometric uncertainty. For example, in \cite{Filion2}, the relative photometric error on the magnitude in the F606W and F814W bands is of the order of $\sim0.2\%$. To take these uncertainties into account --- the photometric one, but also the extra scatter due to the non-exactitude of the isochrone hypothesis --- we generate a sample similar to the one used for the main study of this paper, but with a Gaussian noise of $1\%$ on the F606W and F814W magnitudes of each star. The amplitude of the Gaussian noise is chosen in such a way that the resulting colour-magnitude diagram is similar to real diagrams obtained from observational studies (cf. \citealt{Geha,Gennaro1,Gennaro2,Filion2}). This colour-magnitude diagram is shown on Fig.~\ref{fig:CMD_UncertPop}.

\begin{figure*}
\includegraphics[width=1.\columnwidth]{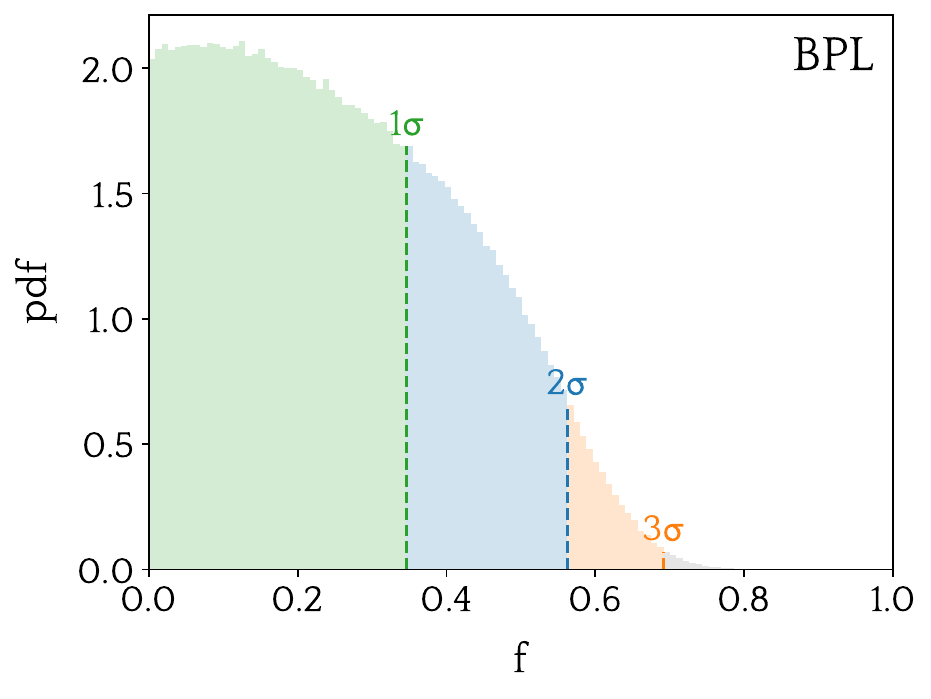}
\includegraphics[width=1.\columnwidth]{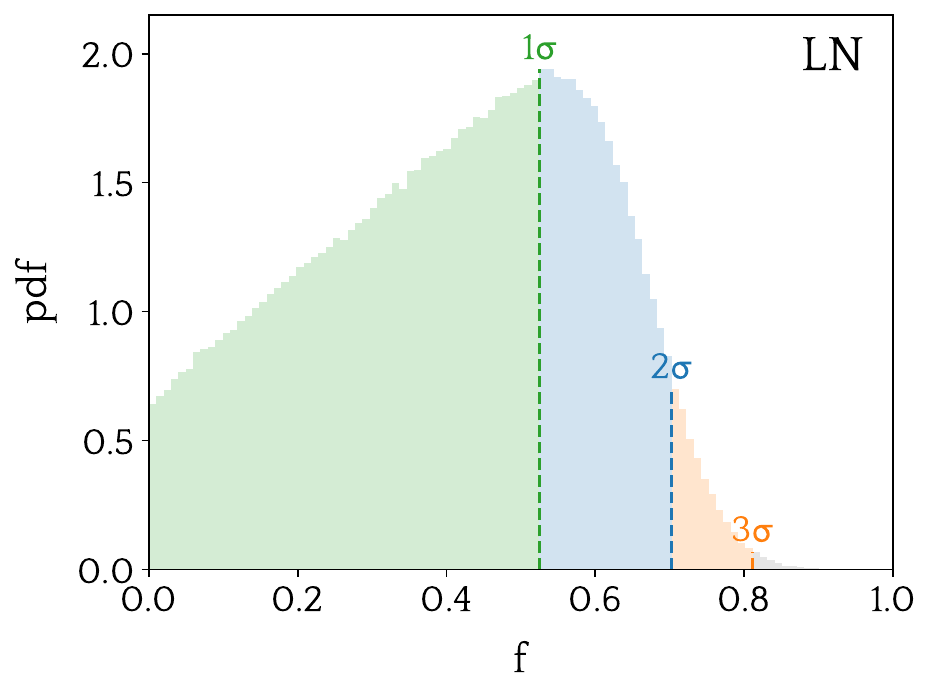}
\includegraphics[width=1.\columnwidth]{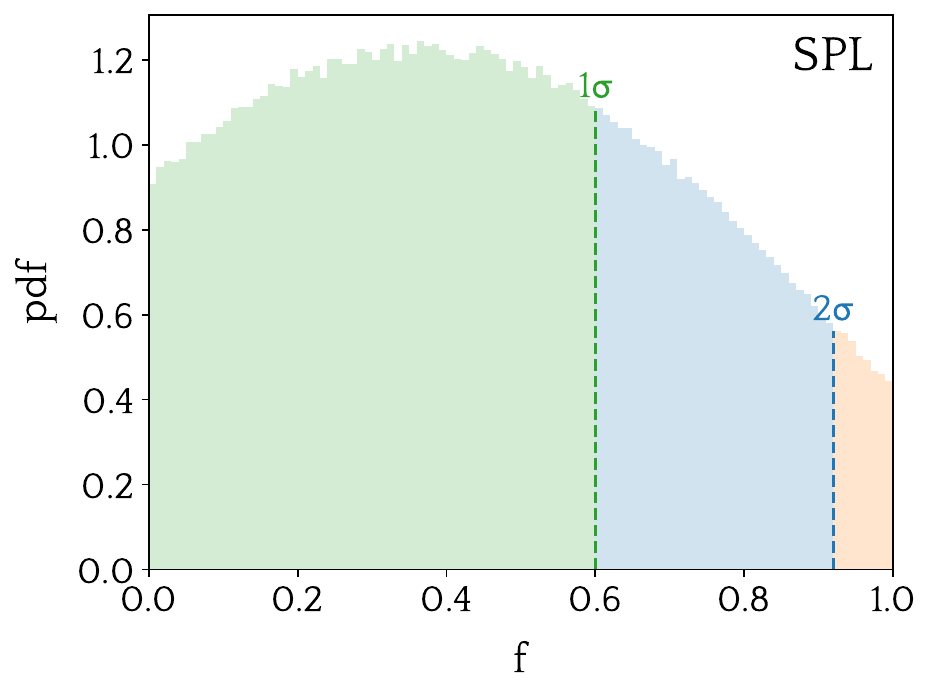}
\caption{\label{fig:BPLandLNandSPLintegrated_UncertPop}
Marginalized probability density function of the PBH fraction $f$ for the BPL (top left panel), LN (top right panel) and SPL (bottom panel) cases with the scattered sample. The left (green) region corresponds to the values that are within one-sided $1\sigma$, the middle (blue) region corresponds to points between $1\sigma$ and $2\sigma$ and the right (orange) region corresponds to points between $2\sigma$ and $3\sigma$.}
\end{figure*}

We performed the analysis described in section \ref{sec:BayesAnalysis} on this new sample, and obtained the results depicted in Fig.~\ref{fig:BPLandLNandSPLintegrated_UncertPop}. One can see that the exclusion limits of the fraction $f$ are very close to the original ones. Hence, we conclude that adding scatter to the mock data sample does not change the constraining power of the method significantly. 

\label{appendix}

\bsp
\label{lastpage}
\end{document}